# ELECTROWEAK TOP QUARK PRODUCTION AT THE FERMILAB TEVATRON[*]

A.P. HEINSON

*Department of Physics, University of California,*
*Riverside, CA 92521–0413, USA*

A.S. BELYAEV and E.E. BOOS

*Nuclear Physics Institute, Moscow State University,*
*RU–119899 Moscow, Russia*

hep-ph/9509274  12 Sep 1995

## ABSTRACT

Top quarks can be produced from the electroweak $Wtb$ coupling, leading to a single top or antitop in the final state. We examine single top quark production at the Fermilab Tevatron and calculate the cross section as a function of top quark mass, for each of three separate production processes. We give details of the interferences between Feynman diagrams in $W$-gluon fusion and then study the effects of an anomalous $(V+A)$ term in the $Wtb$ coupling. Using predictions for the next Tevatron run, we estimate the experimental sensitivity to the CKM matrix element $V_{tb}$ as a function of the strength of the anomalous coupling.

## 1. Introduction

The top quark has recently been discovered by the DØ[1] and CDF[2] collaborations at the Fermilab Tevatron. Measurements of its mass are $199 \pm 30$ GeV and $176 \pm 13$ GeV from DØ and CDF in the lepton+jets decay modes and $145 \pm 32$ GeV in the dilepton decay mode from DØ[3]. Because the top quark is so heavy, electroweak production of single top quarks from a $Wtb$ vertex becomes competitive with strong production of $t\bar{t}$ pairs from a $gtt$ vertex. The resummed next-to-leading order cross section for $t\bar{t}$ production[4] is $2.26^{+0.26}_{-0.17}$ pb at $\sqrt{s} = 1.8$ TeV when the top quark mass $m_t = 200$ GeV. A recent calculation to $O(\alpha^2 \alpha_s)$ for single top production[5] gives the cross section for the principal single top process (for $t$ and $\bar{t}$ combined) as 1.7 pb, which is 75% of the total $t\bar{t}$ rate. Therefore, the possibility exists for an experimental measurement of single top production at the Tevatron with the current ~100 pb$^{-1}$ data set.

We have examined the processes for single top production, and calculated the corresponding cross sections, at the current Tevatron energy of $\sqrt{s} = 1.8$ TeV. In addition, we have started to prepare for the large numbers of single top events predicted for the next Tevatron run at $\sqrt{s} = 2.0$ TeV, when there should be twenty times as much data as exists now. We have examined the effects of an anomalous right-handed $(V+A)$ coupling at the $Wtb$ vertex from which single top quarks are produced. The single top quark system is the only place to study this coupling, and to measure the CKM matrix element $V_{tb}$.

There is an extensive literature on single top physics at hadron colliders[5,6], although few papers use the current Tevatron energy with modern parton distributions. How to study the $Wtb$ coupling and what can be learnt have also featured in the literature[7].

---

## 2. Single Top Quark Cross Section

### 2.1 Single Top Processes

We have calculated the tree level cross sections for the following processes:

1. $p\bar{p} \to t\bar{b} + X$
2. $p\bar{p} \to tq + X$
3. $p\bar{p} \to tW^- + X$

where *X* represents additional final state particles other than gluons. In process 1. the final state $t$ and $\bar{b}$ are produced from an extremely off-shell s-channel *W* boson. Process 2. occurs via a t-channel *W* boson. Process three has several types of Feynman diagrams. We have included the following subprocesses in our calculations:

| | | | |
|---|---|---|---|
| 1.1 $q'\bar{q} \to t\bar{b}$ | 1.2 $q'g \to t\bar{b}q$ | | |
| 2.1 $q'b \to tq$ | 2.2 $q'g \to tq\bar{b}$ | | |
| 3.1 $bg \to tW$ | 3.2 $q\bar{q} \to tW\bar{b}$ | 3.3 $gg \to tW\bar{b}$ | |

where *q* is a valence or sea *u* or *d* quark. Subprocess 2.2 $q'g \to tq\bar{b}$ is known as *W*–gluon fusion. Although 1.2 and 2.2 look superficially the same, in fact they each have two Feynman diagrams which form separate gauge invariant sets, and the diagrams are calculated in pairs as higher order corrections to the 2→2 processes with $t\bar{b}$ and $tq$ in the final state. Feynman diagrams for these processes are shown in Fig. 1.

In our calculations, we have omitted subprocesses with strange or charm sea quarks in the initial state for simplicity. These would boost the total cross section by ~2–3%. We have included diagrams with intermediate state photons or *Z* bosons where appropriate (3.2 and 3.3) but the contributions are not significant. For subprocesses 3.2 and 3.3, we have omitted diagrams with strong $t\bar{t}$ production, where the $\bar{t}$ has decayed to a $W\bar{b}$. We have included diagrams with off-diagonal CKM matrix elements.

The subprocesses we have included in our calculations comprise all the significant ones with two or three vertices, except those with a gluon in the final state, which are significant but have not yet been fully treated. Subprocesses with an extra quark in the final state, for instance $bg \to tq\bar{q}'$ and $q\bar{b} \to tWq$, although they have several Feynman diagrams, do not contribute more than ~1–2% to the total cross section. This also applies to the subprocess $b\bar{b} \to tW\bar{b}$, despite its large number of Feynman diagrams, including one with electroweak $t\bar{t}$ production.

### 2.2 Calculation Details

We have calculated the production cross section for each of the subprocesses mentioned above. We used the computer program CompHEP[8] to do the tree level symbolic calculations and to generate optimized FORTRAN code for the squared matrix elements. We used the BASES[9] package to integrate over all phase space using parton distribution functions, and a CompHEP–BASES interface[10] to generate the event kinematics with smoothing of singular variables. For parton distributions, we have used the CTEQ3M[11] and MRS($A'$)[12] fits, which are representative next-to-leading-order sets in the $\overline{MS}$ renormalization scheme.

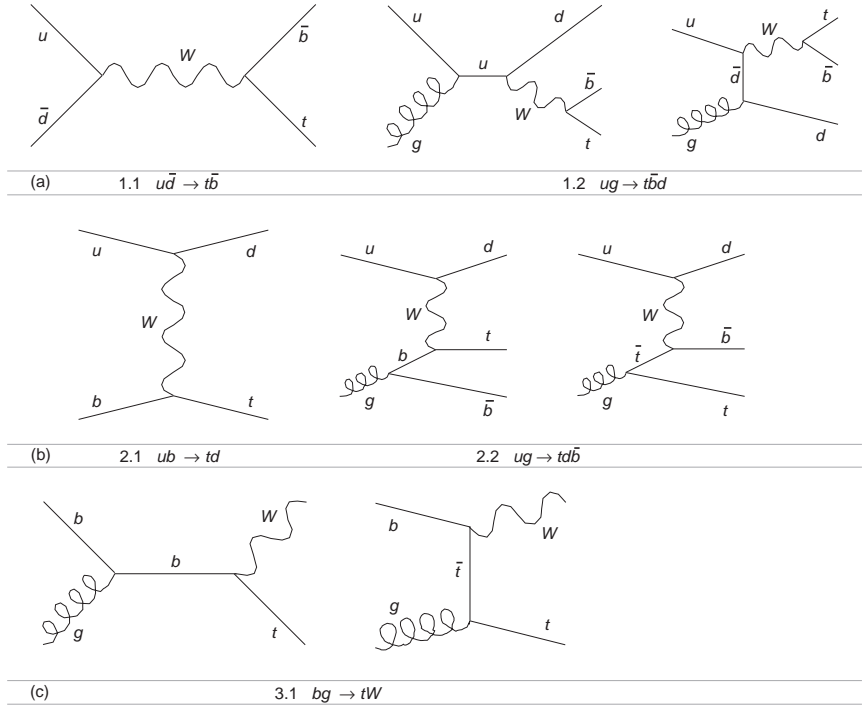

Fig.1 Representative Feynman diagrams for the significant subprocesses included in our calculation for single top quark production at the Tevatron: (a) $W$ boson s-channel production $p\bar{p} \to t\bar{b} + X$; (b) $W$ boson t-channel production $p\bar{p} \to tq + X$ and (c) $p\bar{p} \to tW^- + X$.

The following Standard Model parameters have been used in our calculations: $m_Z$ = 91.19 GeV, $m_b$ = 5.0 GeV, $\alpha$ = 1/128, $\sin^2 \vartheta_W$ = 0.225 and CKM matrix elements $V_{ud}$ = 0.975 and $V_{tb}$ = 0.999. All results have been obtained in the unitarity gauge and the 't Hooft–Feynman gauge, as a check of the calculations. Differences between calculations in the two gauges are less than 0.1%.

We have chosen to use $m_t^2$ as the QCD evolution parameter or scale $Q^2$. A typical $x$ value is 0.1, where $x$ is the fraction of the proton or antiproton momentum carried by each initial state parton. At a scale $Q^2 = (180 \text{ GeV})^2$, the value of $\alpha_s$ is 0.104.

*2.3 Combining Cross Sections*

Care must be taken when combining some single top subprocesses to avoid double counting. Subprocesses 2.1 ($q'b \to tq$) and 2.2 ($q'g \to tq\bar{b}$: $W$-gluon fusion) have a considerable region of overlap when the $g \to b\bar{b}$ is on-shell, so that the $b$ quark which fuses with the $W$ boson is the same as the initial state $b$ sea quark in subprocess 2.1. A technique has been developed[13] which takes care of this double counting whilst being consistent with the definition of the $b$ sea quark in the parton distributions. It involves summing the cross sections calculated for each of the two subprocesses, and then subtracting the gluon splitting rate convoluted with subprocess 2.1. We have adopted this technique here.

The subtracted term is not a small correction. For example, using CTEQ3M and $m_t$ = 180 GeV, the 2→2 subprocess $q'b \to tq$ gives a cross section of 0.750 pb before subtraction, $W$-gluon fusion $q'g \to tq\bar{b}$ gives 0.293 pb and the correction is –0.536 pb, giving a total rate for $p\bar{p} \to tq + X$ of 0.507 pb. The subtraction term forms just over half of the raw total rate. Alternatively, if the 2→2 subprocess and the subtraction term are just ignored, then the remaining cross section from only the $W$-gluon fusion subprocess is less than 60% of the total $p\bar{p} \to tq + X$ rate.

*2.4  Cross Section Versus Top Quark Mass*

Our results are shown in Fig. 2 as a function of the top quark mass $m_t$, at $\sqrt{s}$ = 1.8 TeV. The solid curves are from the calculations with CTEQ3M and the dot-dash curves are from MRS($A'$). With $m_t$ = 180 GeV, we find the total cross section for top to be 0.834 pb (CTEQ3M), with contributions of 0.247 pb from subprocess 1.1, 0.032 pb from 1.2, 0.215 pb from 2.1 after correction, 0.293 pb from 2.2 ($W$-gluon fusion), 0.042 pb from 3.1, and 0.005 from subprocesses 3.2 and 3.3 combined.

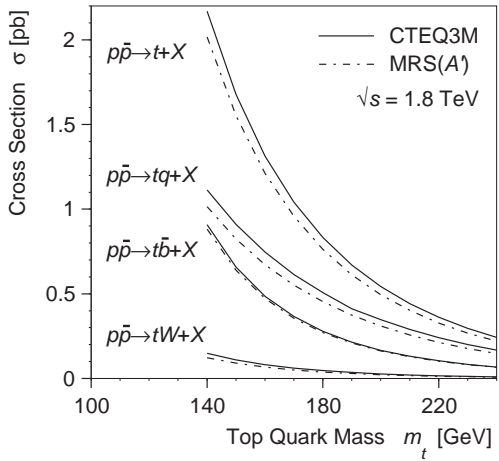
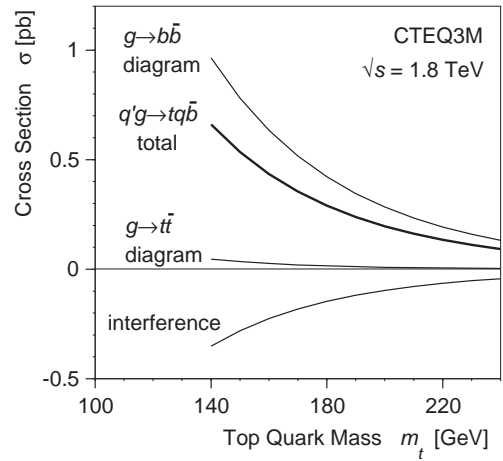

Fig.2  Electroweak single top quark production cross section from $p\bar{p}$ interactions.

Fig.3  Interference between the Feynman diagrams of the $W$-gluon fusion subprocess $q'g \to tqb$.

## 3.  A Closer Look at *W*-Gluon Fusion

We have studied the contributions to the production rate from the two Feynman diagrams in subprocess 2.2, $W$-gluon fusion: $q'g \to tq\bar{b}$. These calculations were done in the unitarity gauge with the CTEQ3M parton distributions. The Feynman diagram in which the initial state gluon produces a $b\bar{b}$ pair contributes 145% to the total rate, whereas the diagram in which the gluon splits to a $t\bar{t}$ pair contributes only 5%. Although this diagram contributes so little to the total, the destructive interference between the two diagrams is very significant, at –50%. Figure 3 shows the contributions to the total *W*-gluon fusion cross section from each diagram and from the interference between them as a function of the top quark mass.

## 4. *Wtb* Coupling and $V_{tb}$

We have investigated the effect of an anomalous contribution to the *Wtb* coupling, in the form of a right-handed (*V+A*) structure, specified by a parameter $A_r$. In the unitarity gauge, the *Wtb* coupling is given by:

$$\Gamma = \frac{eV_{tb}}{2\sqrt{2}\sin\vartheta_W}\left[\gamma_\mu(1-\gamma_5) + A_r\gamma_\mu(1+\gamma_5)\right]$$

where *e* is the electronic charge, $\sin\vartheta_W = 0.474$ and $\gamma_\mu$ and $\gamma_5$ are Dirac matrices.

The dependence of the total single top cross section on the parameter $A_r$ is shown in Fig. 4 for the upgraded Tevatron energy of $\sqrt{s} = 2.0$ TeV and $V_{tb} = 0.999$. Here $\sigma(t+\bar{t}) = \sigma\left((t\bar{b}) + (\bar{t}b) + (tq) + (\bar{t}\bar{q}) + (tW^-) + (\bar{t}W^+)\right)$. The cross section rises from the SM value of 2.37 pb to 4.60 pb when $A_r = +1$.

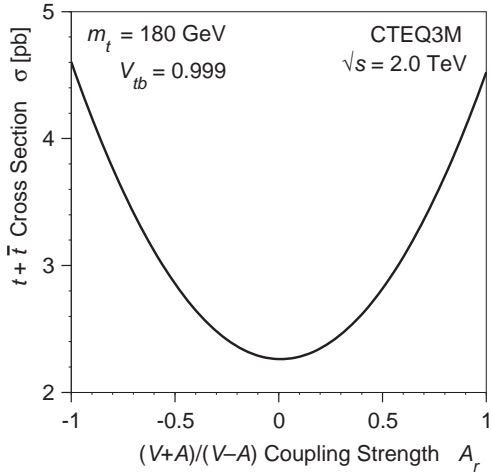 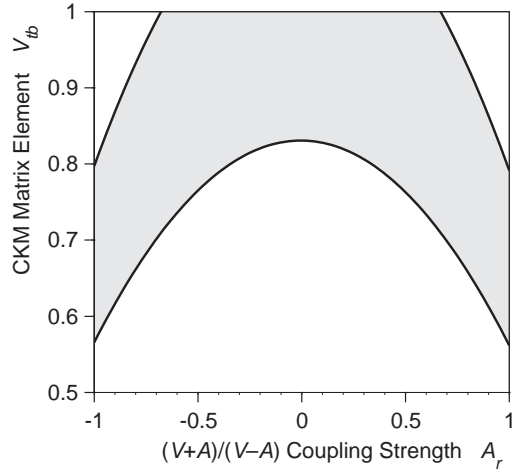

Fig.4 Total single top quark production cross section in $p\bar{p}$ interactions as a function of the right-handed (*V+A*) coupling parameter $A_r$.

Fig.5 Region of sensitivity at the 95% C.L. in the ($V_{tb}, A_r$) plane, for an experiment with 2 fb$^{-1}$ of data, S:B = 1:2, 10% efficiency and $m_t$ = 180 GeV.

We have estimated the experimental sensitivity in the ($V_{tb}, A_r$) plane from the next Tevatron collider run. The parameters used in our estimate are that the collider will operate at 2.0 TeV, there will be a 2 fb$^{-1}$ data set, the top quark has a mass of 180 GeV, and the search will take place in the electron+jets and muon+jets decay modes with 10% overall efficiency. We assume the branching fraction will remain constant at 2/9 and that the signal-to-background ratio will be 1:2, based on a study for the TeV-2000 project[14]. If the *Wtb* coupling has SM form and there are only three quark families (i.e. $0.9988 \le V_{tb} \le 0.9995$ at the 90% CL[15]), then there should be approximately 105 fully reconstructed *b* tagged events. Assuming that 315 events are seen (105 signal and 210 background), limits can be placed on $V_{tb}$ as a function of any anomalous coupling as parametrized by $A_r$. These 95% CL limits are shown in Fig. 5. If the $W_{tb}$ coupling is purely left-handed, then the limits on $V_{tb}$ are $0.817 \leq V_{tb} \leq 1$. For $A_r = 0.5$, $0.735 \leq V_{tb} \leq 1$, and if $A_r = 1.0$, then $0.586 \leq V_{tb} \leq 0.826$.

## 5. Acknowledgements

We would like to thank P. Baringer, P. Ermolov, P. Grannis, A. Klatchko, W.-K. Tung, S. Willenbrock and C.-P. Yuan for useful discussions, and the Moscow State University CompHEP group, especially S. Ilyin and A. Pukhov for their valuable help. This paper is an abridged version of a more detailed one in preparation[16].